\documentclass[aps,prl,twocolumn,superscriptaddress]{revtex4-1}  
\usepackage{graphicx}  % needed for figures % http://ctan.org/pkg/graphicx
\usepackage{dcolumn}   % needed for some tables
\usepackage{bm}        % for math
\usepackage{amssymb}   % for math
\usepackage{siunitx}
\usepackage[breaklinks=true,colorlinks,citecolor=blue,linkcolor=red,urlcolor=blue]{hyperref}
\usepackage{multirow}
\usepackage{xcolor}
\usepackage{amsmath}
\usepackage{hhline}
\usepackage{lipsum}% http://ctan.org/pkg/lipsum

\usepackage{soul}
\usepackage{color} 

\usepackage{braket}
\usepackage{autobreak}

\usepackage{amsfonts}
\usepackage{amsmath}
\usepackage{slashed}
\usepackage{amsthm}
\usepackage{amssymb}

\newcommand{\Z}{\mathbb{Z} }

\begin{document}

\title{Topological Orbital Hall Effect}

%% Authors %%%

\author{Baokai Wang}
\email{wang.ba@northeastern.edu}
\thanks{B. W. and Y.-C. H. contributed equally to this work.}
\affiliation{Zhejiang Institute of Photoelectronics, Zhejiang Normal University, Jinhua, Zhejiang 321004, China}
\affiliation{Research Center for Industries of the Future $\mathcal{\&}$ Key Laboratory for Quantum Materials of Zhejiang Province, Department of Physics, Westlake University, Hangzhou, Zhejiang 310024, China}
\affiliation{Department of Physics, Northeastern University, Boston, Massachusetts 02115, USA}
\affiliation{Quantum Materials and Sensing Institute, Northeastern University, Burlington, MA 01803, USA}

\author{Yi-Chun Hung}
\email{hung.yi@northeastern.edu}
\thanks{B. W. and Y.-C. H. contributed equally to this work.}
\affiliation{Department of Physics, Northeastern University, Boston, Massachusetts 02115, USA}
\affiliation{Quantum Materials and Sensing Institute, Northeastern University, Burlington, MA 01803, USA}

\author{Hsin Lin}
\affiliation{Institute of Physics, Academia Sinica, Taipei 115201, Taiwan}

\author{Sheng Li}
\affiliation{Zhejiang Institute of Photoelectronics, Zhejiang Normal University, Jinhua, Zhejiang 321004, China}

\author{Rui-Hua He}
\affiliation{Research Center for Industries of the Future $\mathcal{\&}$ Key Laboratory for Quantum Materials of Zhejiang Province, Department of Physics, Westlake University, Hangzhou, Zhejiang 310024, China}

\author{Arun Bansil}
\email{ar.bansil@northeastern.edu}
\affiliation{Department of Physics, Northeastern University, Boston, Massachusetts 02115, USA}
\affiliation{Quantum Materials and Sensing Institute, Northeastern University, Burlington, MA 01803, USA}

%\date{\today}

\begin{abstract}
The orbital Hall effect (OHE) is attracting recent interest due to its fundamental science implications and potential applications in orbitronics and spintronics. Unlike the spin Hall effect, the connection between the OHE and band topology is not well understood. Here we present a novel approach for understanding the OHE based on analyzing the projected orbital angular momentum (POAM) spectrum. By considering monolayers of group IV elements, we demonstrate that the Wannier charge centers of the POAM spectrum display topologically nontrivial windings. The orbital Hall conductivity is found to form a plateau within the band gap as a direct consequence of the Chern number carried by the POAM spectrum. The topological orbital Hall phase is shown to yield a new form of bulk-boundary correspondence, which features gapless states in the POAM spectrum and induces nonzero orbital textures at the boundaries that should be amenable to experimental verification through ARPES measurements. Our study presents a systematic method for investigating the topological OHE and provides a pathway for its broader exploration in two-dimensional materials.

\end{abstract}

\maketitle

% ==============================================.

Recent advances in understanding the quantum spin Hall effect (QSHE) indicate that the topology of the ground state and the associated topological edge states are of key importance for developing spintronics applications of the QSHE \cite{Zhang_2015, Xiu2011, PhysRevX.1.021001, C7CS00125H, doi:10.1126/science.1174736}. The non-trivial topology here is characterized by a (nearly) quantized spin Hall conductivity (SHC) \cite{TI_1, TI_2, TI_3, Wang_2024} and a spin-polarized edge current \cite{qsh_graphene, qsh_konig, qsh_zhang}. In contrast to spintronics, the spin-orbit torque in orbitronics is generated by leveraging the orbital-angular momentum (OAM), where the orbital Hall effect (OHE) plays a  fundamental role \cite{Go_2021, Choi2023, PhysRevLett.95.066601, PhysRevLett.123.236403, PhysRevMaterials.6.095001, PhysRevB.106.024410, PhysRevB.107.134423, doi:10.1021/acs.nanolett.4c00430, PhysRevMaterials.6.074004, PhysRevLett.128.176601, PhysRevB.108.245105, PhysRevResearch.4.033037, PhysRevB.108.L180404, PhysRevResearch.5.043052, PhysRevResearch.5.043294, PhysRevLett.121.086602, PhysRevResearch.2.013127, Lee2021_1, Lee2021_2}. Unlike the QSHE, which has been shown to be linked closely to the ground-state topology, the potential connection between the OHE and band topology is not clear and remains to be explored \cite{PhysRevB.110.085412, Chen2024, tohe_sky}.

Recent work based on analyzing the topology of spectral features provides a systematic approach for understanding the topology associated with the orbital degree of freedom \cite{feature, PhysRevB.109.155143}. This involves considering the topology of the projected spectrum $\langle P\hat{O}P \rangle$, where $P$ denotes the projection operator for the relevant Hilbert subspace and $\hat{O}$ represents the feature operator of interest that respects the translational symmetry. The feature-dressed Bloch states can demonstrate nontrivial windings within the Brillouin zone (BZ), leading to unique topological characteristics. In this framework, the QSHE is recognized by non-trivial Chern numbers in two spin sectors of the spin spectrum using the feature operator as spin $\hat{O}=\hat{n}\cdot\hat{S}$ with  $\hat{n}=\hat{z}$ \cite{feature, Lin2024, PhysRevB.80.125327, shulman2010robust, PhysRevB.109.155143, Wang_2024, PhysRevB.108.245103}. Similarly, the OHE can be explored by selecting $\hat{O}=\hat{L}_z$, where $\hat{L}$ is the OAM operator.

Monolayers of Group IV elements were among the first candidate materials used for investigating quantum spin Hall insulators, where the existing literature is focused mainly on the QSHE with paucity of work on the OHE. With this motivation, we address the OHE in this study using the feature-spectrum topology approach. We show that the valence states of group IV monolayer films can be divided into three sectors based on the feature spectrum $\langle P\hat{L}_zP \rangle$, which host nonzero Chern numbers, leading to the bulk-boundary correspondence in the OAM spectrum on the edges of the films. We attribute the in-gap plateau observed in the orbital Hall conductivity (OHC) and its distribution within the Brillouin zone (BZ) to the Chern numbers and Berry curvature in the $\hat{L}_z$-spectrum, arising from band inversion in the occupied subspace. We show that the bulk-boundary correspondence in the OHE phase results in nonzero orbital textures at the edges, which should be amenable to experimental verification via  ARPES measurements \cite{PhysRevLett.132.196401}. Since the OHE in the Group IV films has a topological origin, we will refer to such an OHE as a \emph{topological OHE}.

\begin{figure}[ht]
\includegraphics[width=8.5cm]{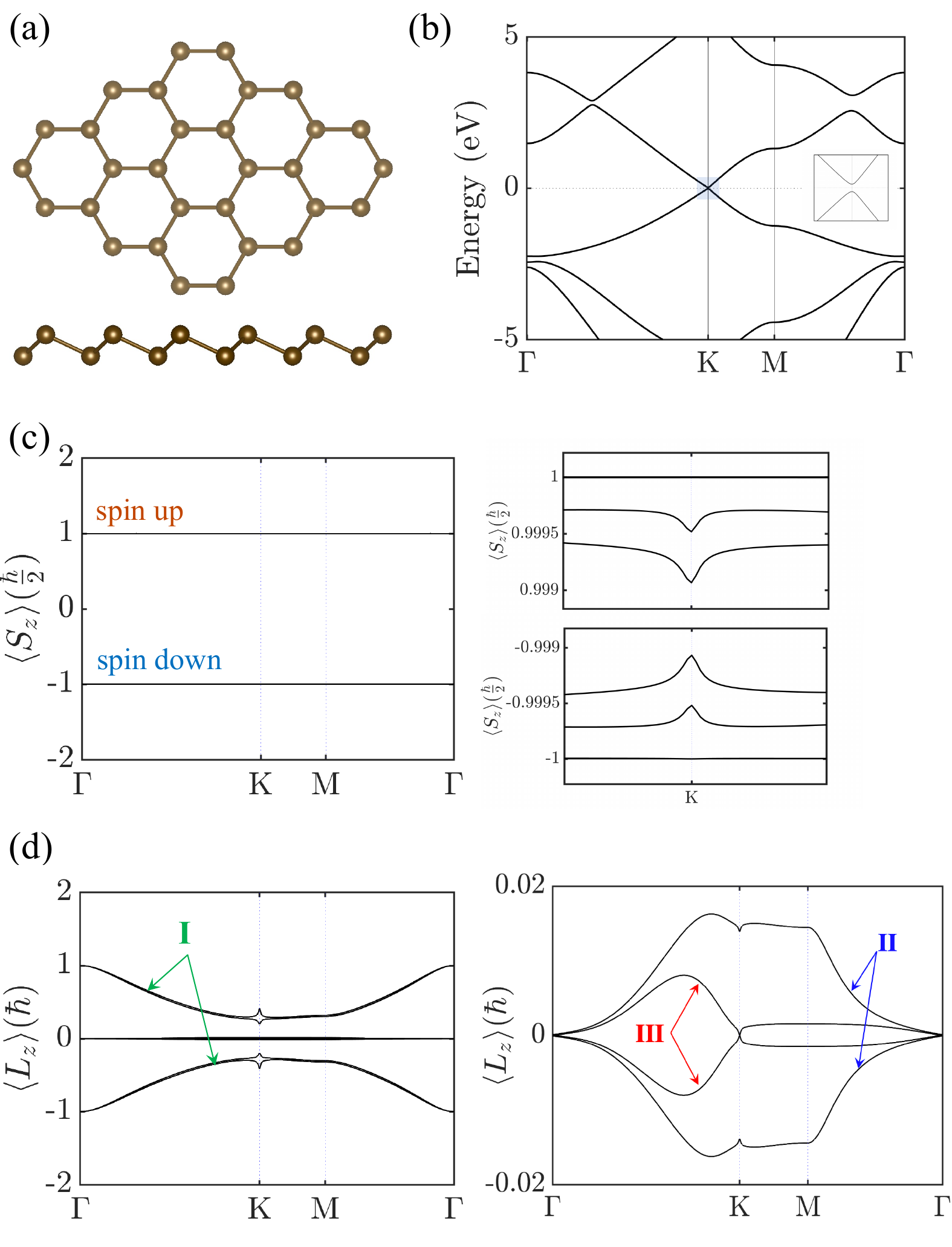}
\caption{(a) Crystal structure of monolayers Group IV elements. The top and bottom panels are the top and side views, respectively. (b) Two-dimensional band structure of monolayer germanene. Inset gives a zoomed-in view of the shaded region concentrated around the K points with a band gap near the Fermi energy. (c) Spectrum of the $\hat{z}$ component of spin, $\langle PS_zP\rangle$. Two right panels give the zoomed-in view of the spectrum around the K point for $S_z = \pm1$ bands. (d) Spectrum of the $z$-component OAM, $\langle PL_zP\rangle$. Right panel gives a close-up view of the bands near $L_z=0$. Three distinct sectors are labelled as I, II, and III.
  }
\label{fig1}
\end{figure}

% ==============================================.
\paragraph*{Electronic structures---} The family of monolayers of Group IV elements  crystallizes in a honeycomb lattice with two sublattices misaligned in the out-of-plane direction, forming a vertical buckling that increases with the atomic mass, see FIG.~\ref{fig1}(a). Without loss of generality, in the following discussion, we take the monolayer germanene as the representative of this family due to its relatively large spin-orbit coupling (SOC). All results presented in this paper are based on a tight-binding model\cite{model}, which faithfully captures the band structure around the Fermi level. Monolayers of group IV elements share similar band structures. The most prominent feature in the band structure is the presence of Dirac cones at the corners of the BZ (FIG.~\ref{fig1}(b)). The Dirac cones are gapped by spin-orbit coupling (SOC) effects, where the magnitude of the gap increases with atomic mass and the degree of buckling (inset of FIG.~\ref{fig1}(b)). 

\begin{figure}[ht]
\includegraphics[width=8.5cm]{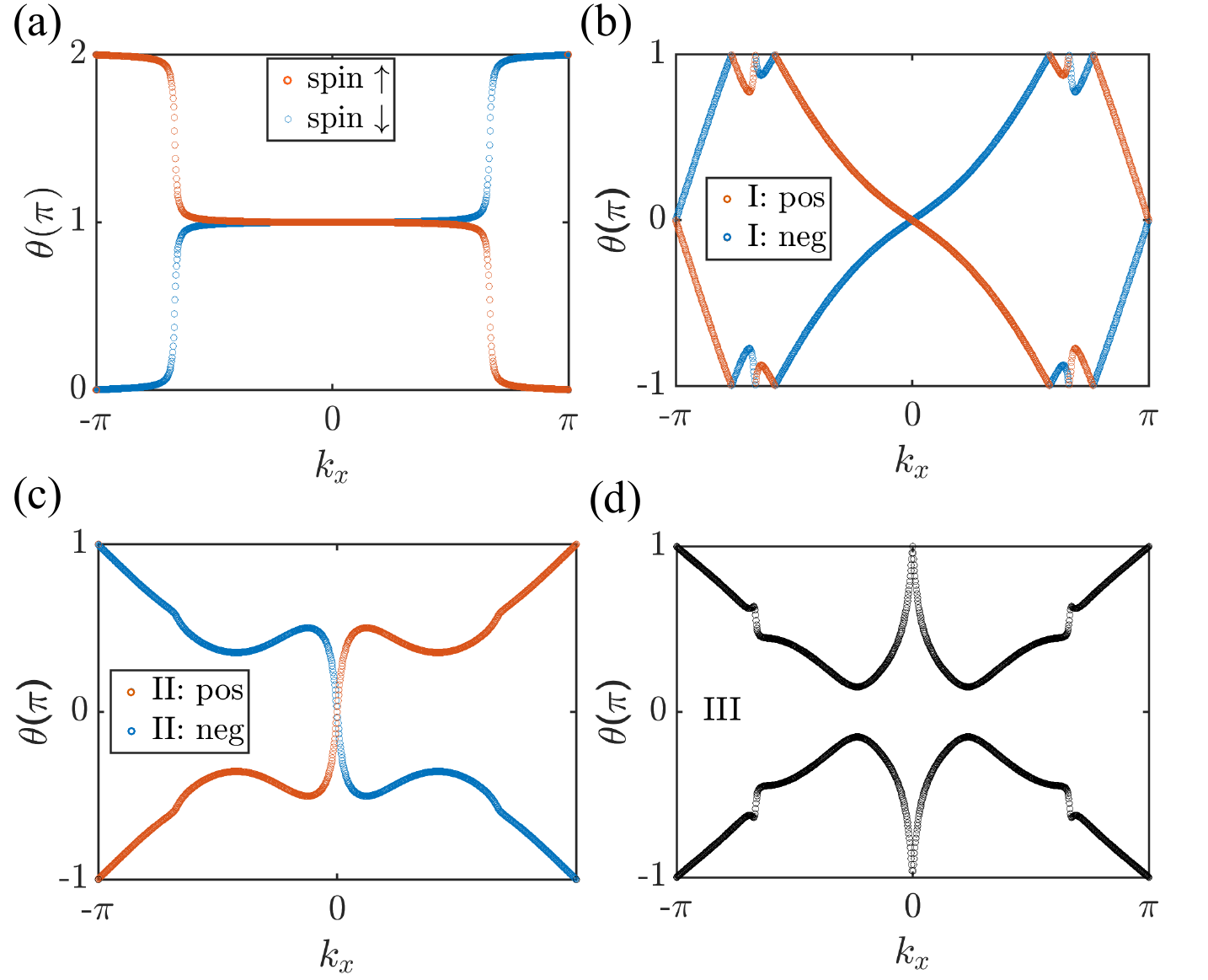}
\caption{(a) Spectrum of the Wannier charge centers for the spin sectors corresponding to $s_z=\pm\frac{\hbar}{2}$. (b-d) Winding of the Wannier charge centers for the OAM sectors I,II, and III.
  }
\label{fig2}
\end{figure}

% ==============================================.
\paragraph*{Feature-spectrum topology---}The feature spectrum $\langle P\hat{O}P\rangle$ partitions the occupied manifold into distinct sectors according to spectral values, where various sectors consist of groups of isolated bands in the feature spectrum. These sectors form subspaces in the occupied manifold, which can exhibit nontrivial topological properties reflected in the nonzero winding of their Wannier charge centers (WCCs) \cite{wcc} calculated from their dressed-Bloch states \cite{feature}. In the case of a quantum spin Hall insulator, the spin spectrum is divided into two (spin up and spin down) sectors $\phi_{\pm}$. The spin-dressed Bloch states are given by $\Phi_{\pm} = \phi_{\pm} * \psi_{nk}$, from which the topological properties can be calculated. As an example. the spin spectrum of monolayer germanene is shown in FIG.~\ref{fig1}(c). Although the two spin sectors appear nearly flat due to the small SOC in group IV elements, a closer inspection reveals that these sectors display a spike feature at the K point (inset of FIG.~\ref{fig1}(c)), which indicates a spin spectrum inversion due to the SOC, suggesting a potential topologically nontrivial nature of the spin spectrum. This echoes the winding of the WCC for each spin sector of the valence bands in germanene (FIG.~\ref{fig2}(a)), which reflects the Chern number for each spin sector $\Phi_{\pm}$ is $\mathcal{C}_{\pm}=\mp 1$. Consequently, the spin Chern number is $\mathcal{C}_s = \frac{1}{2}(\mathcal{C}_{+} - \mathcal{C}_{-}) = -1$.

We now turn our attention to the OHE. Within the tight-binding model, we adapt the $z$-component of the OAM operator in the basis of $(\ket{s},\ket{p_x},\ket{p_y},\ket{p_z})$ atomic orbitals:
\begin{equation}
\hat{l}_z = i\hbar\begin{pmatrix}
 0 & 0 & 0 & 0\\
 0 & 0 &-1 & 0\\
 0 & 1 & 0 & 0\\
 0 & 0 & 0 & 0
\end{pmatrix}.
\end{equation}
The feature operator for describing the OHE is then $\hat{L}_z = \hat{l}_z\otimes \tau_0 \otimes \sigma_0$, where $\tau and \sigma$ act on the sublattice and spin subspaces, respectively. When projected from the valence states, the spectrum of $\langle P\hat{L}_zP \rangle$ consists of four discrete sectors, corresponding to the states of $|L=1, L_z=1\rangle, |L=1, L_z=-1\rangle, |L=1, L_z=0\rangle, |L=0, L_z=0\rangle$. Due to the typical breaking of rotational symmetry and consequently the $U(1)$ symmetry of the OAM in solids, the spectra of $\langle P\hat{L}_zP \rangle$ exhibit dispersion throughout the BZ, see FIG.~\ref{fig1}(d). The feature bands with $L_z = \pm 1$ are seen to be separated from the middle $L_z = 0$ bands. A close-up look also finds that the feature bands with $|L=1, L_z=0\rangle$ are separated from the $|L=0, L_z=0\rangle$ bands. Notably, the spectrum of $\langle P\hat{L}_zP \rangle$ exhibits a dip at the K points, a feature reminiscent of that found in the bands in the spectrum of $\langle P\hat{S}_zP \rangle$. 
 
We classify the isolated sets of bands in the $\langle P\hat{L}_zP \rangle$ spectrum as: $|L=1, L_z=\pm 1\rangle$ (sector I), $|L=0, L_z=0\rangle$ (sector II), and $|L=1, L_z=0\rangle$ (sector III). Sectors I and II are subdivided into positive and negative spectrum components. We define feature-dressed Bloch states as $\Phi_{ik} = \phi_{ik}\psi_{nk}$, where $\phi_{ik}$ and $\psi_{nk}$ are the Bloch states in the feature and the energy bands, respectively, with $i$ denoting various sectors in the feature spectra and $n$ is the band index. Then, we analyze the  Wannier charge center (WCC) spectrum for each sector of the feature-dressed Bloch states $\Phi_{ik}$ to study their band topology. FIGS.~\ref{fig2}(b) and \ref{fig2}(c) show the WCC windings for bands in sectors I and II, which have the winding numbers $\pm2$ and $\pm1$, respectively. Sector III's feature bands are continuous without a gap, prompting the calculation of the $\Z_2$ number for this sector. FIG.~\ref{fig2}(d) presents the WCC spectrum for sector III, which is topologically trivial but shows a dip at K due to band inversion. Nonzero windings of such feature-dressed Bloch states indicate that monolayer group IV films possess  topologically non-trivial orbital domain characteristics beyond the spin domain.

% ==============================================.
\paragraph*{Feature Hall Conductivities---} In the linear response theory, the feature current density flowing along the $\mu$ direction, $\mathcal{J}_\mu^{\hat{O}}$, can be expressed in terms of the feature conductivity tensor by $\mathcal{J}_\mu^{\hat{O}} = \sum_{\nu}\sigma_{\mu\nu}^{\hat{O}}\xi_{\nu}$. Here, $\xi_\nu$ is the $\nu$-component of the applied electric field; $\mu, \nu, \eta$ label the direction in the Cartesian coordinate system. $\hat{O}$ represents the feature operator, such as the angular momentum $\hat{O}_\eta$, where $O=S,L$ denotes the spin or OAM operator depending on the nature of the induced angular momentum that drifts and $\eta$ denotes its polarization. Such a feature conductivity tensor is given by \cite{SM}:
\begin{equation}
\sigma_{\mu\nu}^{\hat{O}_\eta} = \frac{e}{(2\pi)^2}\sum_{n}\int_{BZ}d^2kf_{nk}\Omega_{\mu\nu n}^{\hat{O}_\eta}(k),
\end{equation}
where the band-resolved feature conductivity texture is
\begin{equation}
\Omega_{\mu\nu n}^{\hat{O}_\eta}(k) = 2\hbar\sum_{m\neq n} \text{Im}\left[ \frac{\langle u_{nk}|\hat{j}_{\mu k}^{\hat{O}_\eta}|u_{mk}\rangle\langle u_{mk}|\hat{v}_{\nu k} | u_{nk}\rangle}{(\epsilon_{nk}-\epsilon_{mk} + i 0^{+})^2}\right].
\end{equation}
The $\nu$-component of the velocity is given by $\hat{v}_{\nu k}$ = $\hbar^{-1}\partial H(k)/\partial{k_\nu}$, where $H(k)$ is the Hamiltonian in the $k$-space. $|u_{nk}\rangle$ is the periodic part of the Block eigenstate associated with band energy $\epsilon_{nk}$. $f_{nk}$ is the Fermi-Dirac function. The spin (orbital) angular momentum current operator that flows along the $\mu$-direction with spin (orbital) polarization in the $\eta$-direction is defined by $j_{\mu k}^{O_\eta} = (\hat{O}_\eta v_{\mu, k} + v_{\mu, k}\hat{O}_\eta)/2$ with $\hat{O} = \hat{S}(\hat{L})$. 

\begin{figure}[ht]
\includegraphics[width=8.5cm]{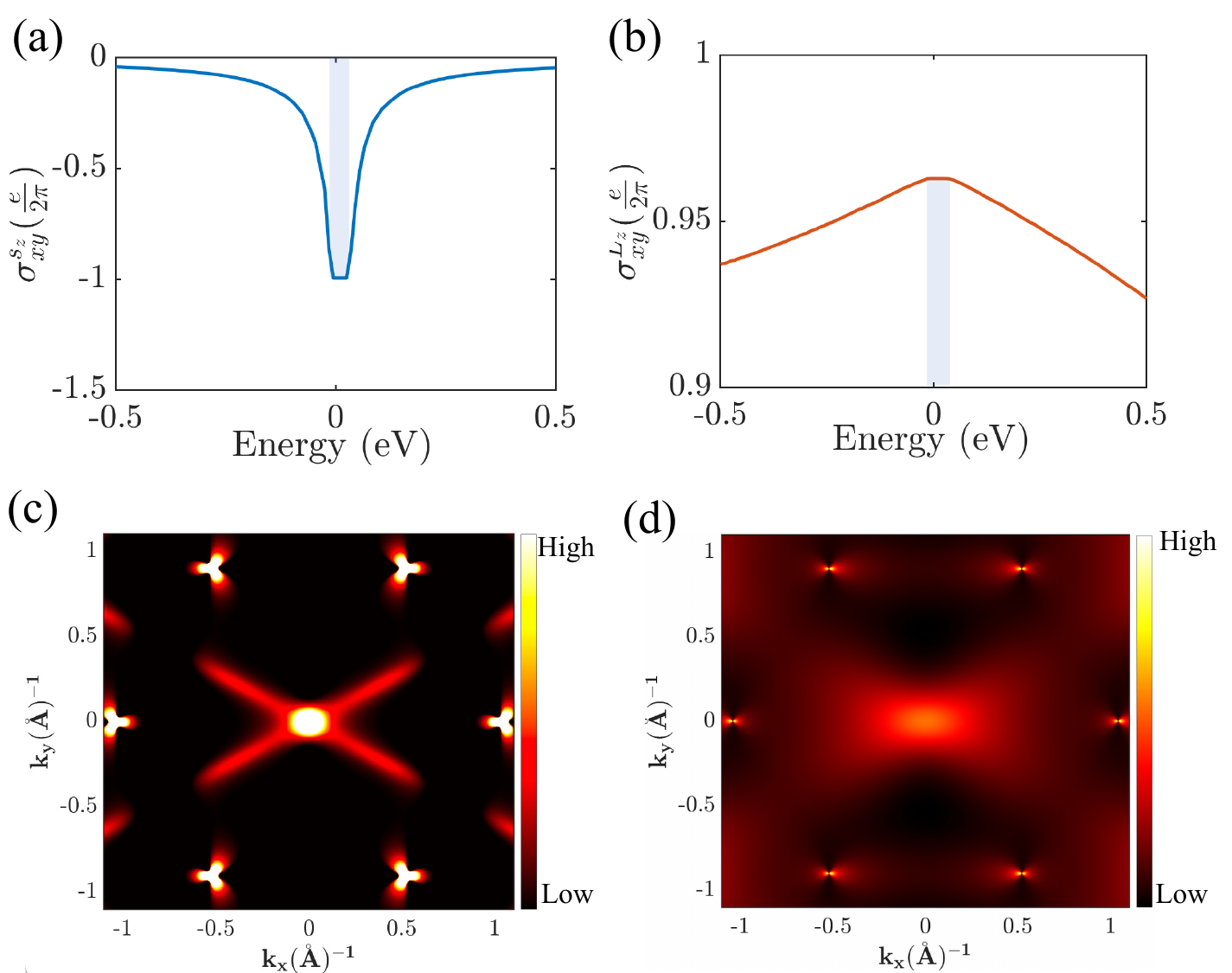}
\caption{(a) Spin and (b) orbital Hall conductivity as a function of the chemical potential in monolayer germanene. The shaded stripe marks the band gap. (c) Distribution of Berry curvature of the bands in the $\hat{L}_z$ feature spectrum in the Brillouin zone. (d) Distribution of orbital Hall conductivity $\sigma_{xy}^{L_z}$ in the Brillouin zone.
  }
\label{fig3}
\end{figure}

FIG.~\ref{fig3}(a) plots the SHC of monolayer germanene, which exhibits a plateau inside the bulk band gap that is almost quantized with $\sigma_{xy}^{\hat{S}_z} = -\frac{e}{2\pi}$. The slight discrepancy between the plateau and the quantized value is due to the weak strength of the spin-$U(1)$ symmetry-breaking SOC. From a mathematical standpoint, this discrepancy can be explained through perturbation theory. When a $U(1)$ symmetry, like $\hat{O}_z$ with $\hat{O} = \hat{S}(\hat{L})$, is chosen to be the feature operator, the feature Hall conductivity is determined by summing the Chern numbers of subspaces in the occupied manifold with distinct eigenvalues of the feature operator: $\sigma_{xy}^{\hat{O}_z} = \frac{e}{h}\sum_{\tilde{o}_i} \tilde{o}_i\mathcal{C}_i$. Here, $\tilde{o}_i$ is the eigenvalue of the feature operator for the $i$th subspace, and $\mathcal{C}_i$ is the corresponding Chern number. This results in a quantized feature Hall conductivity within the bulk band gap when $\tilde{o}_i=\frac{n}{2}\hbar$ for $n\in\mathbb{Z}$ and the Chern numbers have opposite values for subspaces with opposite $\tilde{o}_i$, a phenomenon commonly observed in time-reversal symmetric systems when $\hat{S}_z$ and $\hat{L}_z$ are the chosen feature operators. 

\par In the presence of perturbations that break the $U(1)$-symmetry of the angular momentum, the dispersion of the feature spectrum remains minimal, with spectral values close to the eigenvalues of the angular momentum operator, $\tilde{o}_i$, see FIG.~\ref{fig1}(c) for $\hat{S}_z$. Consequently, the feature Hall conductivity is predominantly determined by the sum of the feature Chern numbers, $\tilde{\mathcal{C}}_i$, weighted by $\tilde{o}_i$, adjusted by perturbation matrix elements, $\delta O_z$, and the leading-order variation in the feature curvature due to the basis transformation from the energy band to the feature band:
\begin{equation}\label{eq:04}
    \sigma_{xy}^{\hat{O}_z} \cong \frac{e}{h}\sum_{\tilde{o}_i} \tilde{o}_i\tilde{\mathcal{C}}_i + \frac{e}{\hbar}\int d\mathbf{[k]} \sum_{m\in\text{occ.}}io_m([\tilde{\Omega},\delta U])_{mm} + \mathcal{O}(\delta O_z).
\end{equation}
Here, $\hat{O}_z=\hat{S}_z$ in germanene, and $o_m$ represents its expectation value for the $m$th energy band. $\tilde{\Omega}$ is the Berry curvature of the bands in the $\hat{O}_z$ feature spectrum, and $\delta U$ is the Hermitian matrix from the basis transformation, see Supplementary Material for details \cite{SM}. Since $\delta O_z$ and $\delta U$ are small perturbations, Eq.~\eqref{eq:04} suggests that the feature Hall conductivity remains nearly quantized when $\hat{O}_z$ symmetry is slightly broken.

On the other hand, the OHC $\sigma_{xy}^{\hat{L}_z}$ also exhibits a plateau within the bulk band gap (FIG.~\ref{fig3}(b)), despite the notable disruption of $\hat{L}_z$ symmetry, which causes the feature spectrum to become dispersive (FIG.~\ref{fig1}(d)). Nevertheless, numerical results reveal that the OHC texture of germanene closely corresponds to the Berry curvature of the bands in the $\hat{L}_z$ feature spectrum, calculated using the Fukui-Hatsugai method \cite{doi:10.1143/JPSJ.74.1674, nakamura2024chernnumberstwodimensionalsystems}. For comparison, we rescaled and subtracted the Berry curvature of bands with positive feature spectral values using those with negative values. As illustrated in FIGS.~\ref{fig3}(c) and (d), the Berry curvature and OHC displays substantial contributions near the $K$ and $K’$ valleys, and the $\Gamma$ point. The Berry curvature pattern at the $\Gamma$ point closely resembles the OHC distribution, indicating a topological origin of the OHC. This finding provides the first evidence of the connection between the linear response of the orbital Hall current and the band geometry in an orbital Hall insulator.

\begin{figure}[ht]
\includegraphics[width=8.5cm]{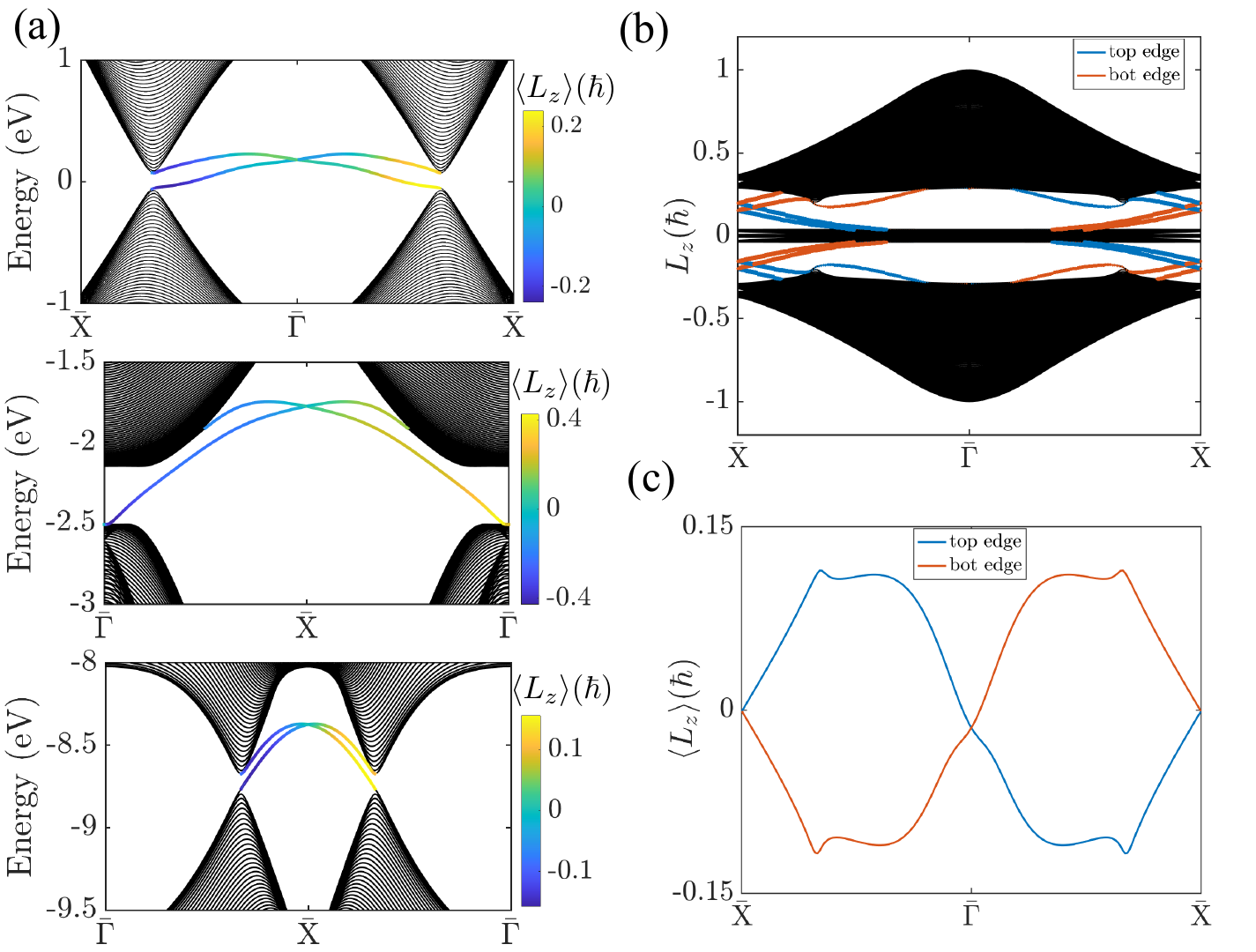}
\caption{(a) Ribbon band structures displaying all the in-gap edge states below the chemical potential along the zigzag direction of monolayer germanene, with the orbital textures of the top edge bands in $\langle l_z\rangle$ using various colors. (b) Edge $\hat{L}_z$ feature spectrum along the zigzag direction, with colors distinguishing the top and bottom edges. (c) Evolution of the expectation value of the OAM of the edge states along both edges of the ribbon.}
\label{fig4}
\end{figure}

% ==============================================.
\paragraph*{Bulk-boundary correspondence---} 
To understand the nature of boundary spectra, we examine the edge band structure of a zigzag ribbon of monolayer germanene. As illustrated in FIG.~\ref{fig4}(a), three sets of edge bands emerge within the valence bands below the Fermi level, each corresponding to a distinct band inversion. The edge bands connecting the valence and conduction bands intersect at $\bar{\Gamma}$, while the Dirac points of the other two sets of edge bands are situated at $\rm{\bar{X}}$. Notably, for all three sets of edge bands, the polarization of OAM reverses its sign across the Dirac point.

We emphasize that even though the OHE exhibits nontrivial topological properties, this does not ensure the presence of gapless edge states because the $U(1)$-symmetry of OAM is broken. In monolayer germanene, the gapless edge states are protected solely by the $\mathbb{Z}_2$ topology. Topological properties of the OHE can be examined through the feature spectrum. FIG.~\ref{fig4}(b) presents the $\langle PL_zP \rangle$ spectrum of the ribbon bands. The positive and negative parts of the bulk-feature-band projections are symmetric due to the TRS. Notably, the two edge bands transverse the gap between sector I and other sectors and align with the Chern numbers of the feature bands in sector I, as discussed above in connection with FIG.~\ref{fig2}. This demonstrates bulk-edge correspondence within the feature domain. 
Since the edge bands in the energy spectrum are gapless at $\bar{\Gamma}$, the feature spectrum at $\bar{\Gamma}$ is not well defined due to the ambiguity in the occupied states at this point \cite{feature}. As a result, the feature spectrum exhibits a sudden jump at $\bar{\Gamma}$, which however is invisible, because it is obscured by the bulk feature bands. To address this irregularity, we apply an in-plane magnetic field, which opens a gap in the edge spectrum to make the occupied states and the corresponding feature spectrum well-defined (FIG. S1 \cite{SM}).

Expectation value of the OAM for all the occupied states equals the sum of the feature eigenvalues of these states in the feature spectrum \cite{feature}. The presence of chiral edge states in the feature spectrum implies a net OAM texture at the boundary. In FIG.~\ref{fig4}(c), we plot the orbital texture at the edges of the ribbon. Two linear regimes centered around $\bar{\Gamma}$ and $\rm{\bar{X}}$ are evident. These two regimes emerge from the edge states in the low-energy band gap and the band gap deep below the chemical potential, see FIG.~\ref{fig4}.

\paragraph*{Discussion---}We note that the nullification of the bulk OAM texture due to $\mathcal{PT}$-symmetry ($\mathcal{P}$ and $\mathcal{T}$ refer to spatial-inversion and time-reversal symmetry, respectively) implies that a non-zero OAM texture at the edges highlights the topological nature of the OHE in germanene. This predicted edge OAM texture should be amenable to experimental verification via ARPES measurements \cite{PhysRevLett.132.196401}. Our methodology is applicable more generally for investigating two-dimensional materials, including phosphorene and transition-metal dichalcogenides \cite{PhysRevB.101.075429, PhysRevB.108.165415, PhysRevB.82.161404, PhysRevLett.130.116204, PhysRevLett.126.056601, PhysRevB.101.161409, PhysRevB.103.195309, PhysRevB.105.195421, PhysRevResearch.6.023271}. Our study provides a systematic framework for exploring the role of topology in the OHE.

% ==============================================.
\paragraph*{Acknowledgement---} 
The work at Westlake University was supported by the National Natural Science Foundation of China (Grants No. 12274353) and National Key R\&D Program of China (Grant No.2022YFA1402200). The work at Northeastern University was supported by the National Science Foundation through NSF-ExpandQISE award No.2329067 and benefited from the resources of Northeastern University’s Advanced Scientific Computation Center, the Discovery Cluster, and the Massachusetts Technology Collaborative. H.L. acknowledges the support by Academia Sinica in Taiwan under grant number AS-iMATE-113-15.

% ==============================================.
\bibliography{ref}
% ==============================================
% ==============================================
\end{document}